\documentclass[conference]{IEEEtran}
\usepackage[utf8]{inputenc}
\usepackage[T1]{fontenc}
\usepackage{microtype}
\usepackage{graphicx}
\usepackage{filecontents}
\usepackage[noadjust]{cite}
\usepackage[hidelinks, bookmarks=false, draft]{hyperref}
\usepackage[keeplastbox]{flushend}
\usepackage{amsmath}
\usepackage{url}
\usepackage{multirow}
\usepackage{dblfloatfix}
\usepackage{mdframed}
\usepackage{enumitem}
\usepackage{amssymb}
\usepackage{subcaption}
\usepackage{caption}
\usepackage{csquotes}

\mdfsetup{skipabove=3pt,skipbelow=3pt}
\begin{document}

\title{Software Professionals are Not Directors:\\ What Constitutes a Good Video?}

\author{\IEEEauthorblockN{Oliver Karras and Kurt Schneider}
	\IEEEauthorblockA{Leibniz Universität Hannover\\ Software Engineering Group\\ 30167 Hannover, Germany\\ Email: \{oliver.karras, kurt.schneider\}@inf.uni-hannover.de}}

\maketitle

\begin{abstract}
Videos are one of the best documentation options for a rich and effective communication. They allow experiencing the overall context of a situation by representing concrete realizations of certain requirements. Despite 35 years of research on integrating videos in requirements engineering (RE), videos are not an established documentation option in terms of RE best practices. Several approaches use videos but omit the details about how to produce them. Software professionals lack knowledge on how to communicate visually with videos since they are not directors. Therefore, they do not necessarily have the required skills neither to produce good videos in general nor to deduce what constitutes a good video for an existing approach. The discipline of video production provides numerous generic guidelines that represent best practices on how to produce a good video with specific characteristics. We propose to analyze this existing know-how to learn what constitutes a good video for visual communication. As a plan of action, we suggest a literature study of video production guidelines. We expect to identify quality characteristics of good videos in order to derive a quality model. Software professionals may use such a quality model for videos as an orientation for planning, shooting, post-processing, and viewing a video. Thus, we want to encourage and enable software professionals to produce good videos at moderate costs, yet sufficient quality.
\end{abstract}


\IEEEpeerreviewmaketitle

\section{Introduction}
Three of the most important goals of requirements engineering (RE) are to create a \textit{clear scope}, \textit{shared understanding}, and high \textit{specification quality} \cite{Fricker.2015c}. Requirements engineers need to achieve these goals to bridge communication gaps between stakeholders and developers \cite{Gulliksen.2003, Callele.}. Such gaps may result in unfulfilled customer expectations due to the communication of incorrect, ambiguous, and non-verifiable requirements \cite{Bjarnason.2011}. 

A \textit{clear scope} refers to a shared vision of the future system. Stakeholders and project members need to share the same system vision to achieve a \textit{shared understanding} \cite{Glinz.2015}. Creighton et al. 
\cite{Creighton.} as well as Antón and Potts \cite{Anton.1998} emphasize the lack of a clear system vision as a key challenge in RE.
The establishment of a shared and clearly defined vision is a challenging task regardless of whether stakeholder and project members meet in person or not \cite{Ambler.2002, Ochodek.2018}. Therefore, successful requirements communication depends on documentation options with high \textit{specification quality} which are suitable for conveying the stakeholders' needs comprehensibly to the project team.

Depending on the development method, there are different possibilities to convey knowledge in RE. One of the most widely used documentation options to convey stakeholders' needs is a written specification as suggested by standards such as ISO/IEC/IEEE $29148$:$2011$ \cite{ISO29148.2011}. However, the use of a written specification for requirements communication is cumbersome since textual documentations including digital versions have the lowest communication richness and effectiveness \cite{Ambler.2002}. The simple handover of a written specification insufficiently supports the rich knowledge transfer that is necessary to develop an acceptable system \cite{Fricker.2010}. Abad et al. \cite{Abad.2016} found the need for a better support of requirements communication that exceeds pictorial representations in written specifications. The authors proposed to invest more efforts in addressing interactive visualizations such as storytelling, for example with videos \cite{Abad.2016}.

The topic of applying videos in RE has been discussed in the recent years and its contributions have been found to be of interesting potential \cite{DeMarco.1990, Carter.2009, Fricker.2015,Karras.2017a}.
In the last $35$ years, several researchers \cite{Feeney.1983, Creighton.2006, Brill.2010, Karras.2016b} proposed approaches for applying videos in RE due to their communication richness and effectiveness \cite{Ambler.2002}. For example, Brill et al. \cite{Brill.2010} demonstrated the benefits of using ad-hoc videos compared to textual use cases in order to clarify requirements with stakeholders. However, Brill et al. clearly stated: ``We give no guidance for creating good videos -- this remains future work'' \cite[p. 2]{Brill.2010}. Many other approaches also use videos but omit the details about how to produce them \cite{Zachos.,Creighton.2006,Karras.2017}. This lack of guidance could be a crucial reason why videos are not an established documentation option in terms of RE best practice \cite{Fricker.2015c}.

Software professionals are not directors. Therefore, they do not necessarily know what constitutes a good video in general and for an existing approach. This lack of skills and knowledge on how to communicate visually with videos impedes the application of video in RE. We assert the following thesis:

\begin{mdframed}
	\begin{itemize}[leftmargin=-2.5mm]
		
		\item[] \textbf{Thesis:}
		If software professionals knew more about the challenges, actual demands, and efforts on how to communicate visually with videos, they could enrich their communication and thus RE abilities.
	\end{itemize}
\end{mdframed}

This paper is structured as follows: In section \ref{sec:video-production-in-requirements-engineering}, we compare existing video production options in RE and take up our position. Section \ref{sec:plan-of-action-a-quality-model-for-videos} presents our proposed plan of action to encourage and enable software professionals to produce good videos on their own. Section \ref{sec:conclusion} concludes the paper.

\begin{table*}[!b]
	\centering
	\captionsetup{justification=centering}
	\caption{Advantages and Disadvantages of the two Video Production Options \\in Requirements Engineering}
	\label{tbl:t1}
	\begin{tabular}{|c|l|l|}
		\hline
		\textbf{Options} & \multicolumn{1}{c|}{\textbf{Outsourcing Video 	Production to Video Professionals}} & \multicolumn{1}{c|}{\textbf{Producing Videos by Software Professionals}} \\ \hline \hline
		\multicolumn{1}{|c|}{\multirow{4}{*}{\textbf{Advantages}}} & High-end videos with enhanced visual impact & No additional communication with others \\ \cline{2-3} 
		\multicolumn{1}{|c|}{} & Work is done by experts & Shorter production time of a video due to internal work \\ \cline{2-3} 
		\multicolumn{1}{|c|}{} & Extensive skills and knowledge & Full control of video production \\ \cline{2-3} 
		\multicolumn{1}{|c|}{} & Professional equipment & Affordable videos \\ \hline \hline
		\multicolumn{1}{|c|}{\multirow{4}{*}{\textbf{Disadvantages}}} & Additional communication with video professionals & Lower quality videos \\ \cline{2-3} 
		\multicolumn{1}{|c|}{} & Longer production time of a video due to external work & Work is done by amateurs \\ \cline{2-3} 
		\multicolumn{1}{|c|}{} & Lower control of video production & Lack of skills and knowledge \\ \cline{2-3}
		\multicolumn{1}{|c|}{} & Expensive videos & Lack of equipment \\ \hline
	\end{tabular}
\end{table*}

\section{Video Production in Requirements Engineering}
\label{sec:video-production-in-requirements-engineering}
The literature mentions two options for producing videos: Either $(1)$ \textit{to outsource the video production to video professionals} \cite{Creighton.2006} or $(2)$ \textit{to have videos produced by software professionals themselves} \cite{Brill.2010}. Both options have their advantages and disadvantages which we briefly discuss in the following. \tablename{ \ref{tbl:t1}} summarizes the advantages and disadvantages of both video production options in RE.

\subsection{Outsourced Video Production by Video Professionals}
The \textit{Software Cinema} approach of Creighton \cite{Creighton.2006} introduced the role of \textit{video producer} which is in charge of shooting video clips. Although the author did not define who fulfills this role, he suggested hiring a specialized film agency and outsource the video production to video professionals.

The outsourcing of video production to video professionals results in high-end videos since the work is done by experts with the necessary knowledge, skills, and equipment. Such high-end videos are more persuasive due to their enhanced visual impact taking emotional aspects into account \cite{Creighton.2006}.

However, outsourcing is expensive and problematic. The external work may cause a longer production time of a video since in-depth communication between software professionals and video professionals is required. This additional indirection creates another communication gap which may cause further misunderstandings that need to be solved. There is also a lower control of the video production which can lead to irrelevant or even useless videos and consequently unnecessary costs, raising the bar for achieving added value \cite{Brill.2010}.

\subsection{Video Production by Software Professionals}
In contrast to Creighton \cite{Creighton.2006}, Brill et al. \cite{Brill.2010} suggested that software professionals produce videos on their own.

According to Brill et al. \cite{Brill.2010} and Broll et al. \cite{Broll.2007}, software professionals can create affordable videos with sufficient quality for purposes in RE. The internal work allows the full control of video production which might cause a shorter production time of a video since no additional communication is necessary. 

Nevertheless, videos produced by software professionals have a lower quality since the work is done by amateurs. Besides professional equipment, software professionals especially lack the skills and knowledge on how to produce good videos.

\subsection{Our Position}
Although both video production options have their advantages and disadvantages (see \tablename{ \ref{tbl:t1}}), we assume that it is easier to counteract the disadvantages of producing videos by software professionals. In the following, we explain the reasons for our position in detail.

\subsubsection{Lower Quality Videos are Not a Problem}
Lower quality videos created with simple equipment, e.g. modern digital video cameras, smartphones, or tablets, are sufficient for purposes of RE \cite{Brill.2010, Broll.2007}. Brill et al. \cite{Brill.2010} demonstrated the benefits of low-effort videos produced by computer scientists who had never applied videos in RE before. Broll et al. \cite{Broll.2007} reported about qualitative lessons learned from applying videos in RE (elicitation, negotiation, validation, documentation). According to their findings, amateur videos created with household equipment are sufficient for RE.

\subsubsection{There is No Need for Better Equipment}
Owens and Millerson \cite{Owens.2011} confirmed the results of Brill et al. \cite{Brill.2010} and Broll et al. \cite{Broll.2007} from a television and video production perspective. The authors stated: ``What was considered a professional quality camera 10 years ago has been dwarfed by the quality of small, low-cost high definition pocket-sized cameras available today. [\dots] So now, no one can blame the lack of quality on his or her camera gear because almost anyone can afford the equipment''\cite[p. 80]{Owens.2011}.

\subsubsection{We Need to Know How to Visually Communicate}
Since the equipment is no longer the main issue of low video quality, Owens and Millerson concluded  ``that the important thing is to know how to visually communicate'' \cite[p. 80]{Owens.2011}. This statement substantiates that a lack of knowledge and skills on how to communicate visually with videos is a crucial concern impeding their successful application. Software professionals are not directors. Thus, they do not necessarily have this required knowledge and skills. Therefore, they need guidance that refers to what constitutes a good video. This guidance does not yet exist. Most existing approaches consider video production as a secondary task and thus neglect the essential step of how to produce videos that are appropriate for their respective purpose.\\

Inspired by the computer science researchers who crossed the boundary of linguistics to increase the impact of written requirements \cite{Mavin.2009}, we came up with the idea of crossing the boundary of video production to increase the impact of video as a medium in RE. Therefore, we decided to learn from the discipline of video production how to visually communicate with videos in order to establish videos as a documentation option in RE.

However, we do not want software professionals to become professional directors. Instead, there is a need to understand the existing know-how of video production to transfer the essential knowledge into RE. We assume the necessity to focus on simplicity with respect to used technology as well as required knowledge and skills so that software professionals can apply videos easier in RE. Therefore, it is necessary to understand what constitutes a good video for successful visual communication. From our point of view, this might be a starting point to establish video as a medium in RE in order to take full advantage of its currently neglected potential.

\section{Plan of Action: A Quality Model for Videos}
\label{sec:plan-of-action-a-quality-model-for-videos}

\subsection{Idea for a Quality Model}
Whether a video is good or not depends on its perceived quality in its respective context of use. A video produced for family and friends may be odd, defocused, lopsided, or with an inadequate perspective, but still be enjoyable for all viewers \cite{Owens.2011}. However, any of the previously mentioned defects are unacceptable to consumers. While a \textit{familiar community}, such as a family, does not expect professional videos with high-end quality, consumers perceive any of such defects as careless, non-\textit{professional} work.

In RE, the context of use of videos often combines aspects of a familiar community and professional work. The involved stakeholders and project members form a small group of mutually known individuals. Thus, they represent a more familiar community collaborating in a professional environment to engineer and build a high-quality system. This combined context impedes the prediction of \textit{how the different viewers (stakeholders and project members) assess the quality of a video} since various quality characteristics affect their attitudes.

Viewers are interested in what a video shows and tells. They are not concerned how the video production was done unless they get bored or the technology becomes obtrusive and distracting \cite{Owens.2011}. Such defects need to be avoided to focus the viewers’ attention on the conveyed content so that they can fulfill the respective goals of their individual underlying information needs. We need to encourage and enable software professionals to produce good videos on their own at moderate costs, yet sufficient quality. Since higher quality usually requires more resources and efforts, there is a need for a suitable balance between sufficient quality and affordable costs.

In consideration of the ISO/IEC FDIS $25010$:$2010$ \cite{ISO25010.2010}, we can deduce that a comprehensive specification and evaluation of the quality of a video requires the definition of the necessary and desired quality characteristics associated with the producers' and viewers' goals and objectives for a video. This definition includes quality characteristics of a video related to its representation, its content, and its impact on the target audience. A quality model for videos following the ISO/IEC FDIS $25010$:$2010$ fulfills this definition and can be used to identify relevant quality characteristics that can be further used to establish requirements, their criteria for satisfaction and corresponding measures for a respective video in RE. However, a video is only a representation format for arbitrary contents that producers want to convey to diverse viewers. Therefore, a quality model for videos always requires the elaboration of its contained quality characteristics in terms of the content and desired impact of a video on its target audience depending on the concrete application in RE. Such an elaborated quality model for videos represents a tailored checklist for ensuring a comprehensive treatment of video quality requirements. Thus, it provides a basis for estimating the consequent efforts and activities that will be needed during the production of a video.

\subsection{Proposed Approach}
\begin{figure}[htbp]
	\centering
	\includegraphics[width=\columnwidth]{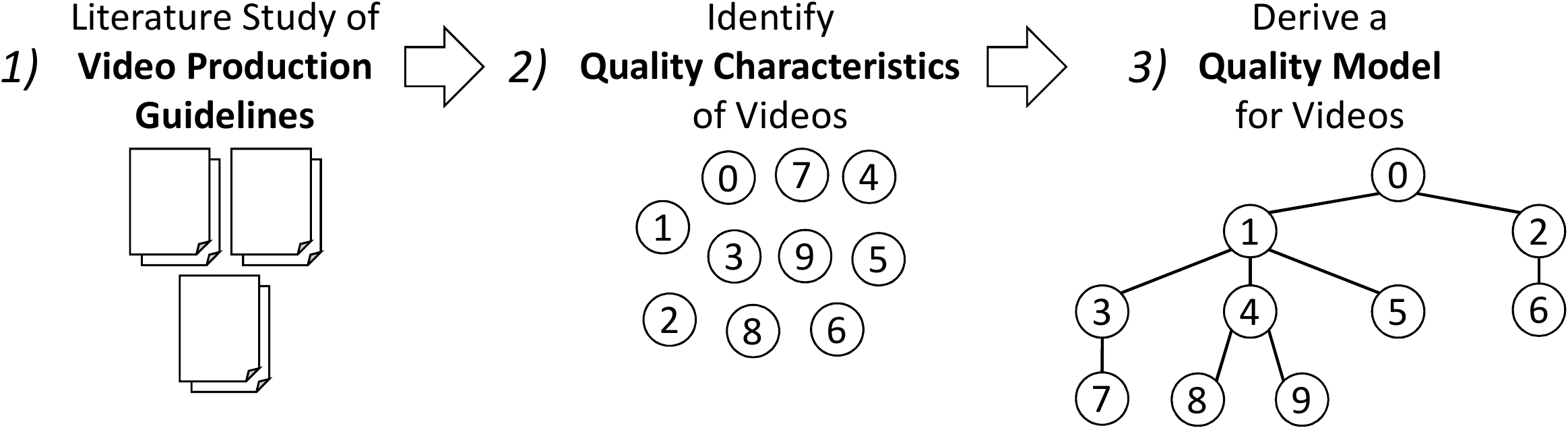}
	\caption{Proposed Approach -- A Quality Model for Videos}
	\label{fig:fig1}
\end{figure}

There are several generic guidelines and recommendations for video production \cite{MIT,Boise,Leeds,ARSC,Owens.2011,Heath.2011}. All of them represent best practice on how to produce good videos with specific characteristics for visual communication. We propose to analyze this existing know-how in order to learn what constitutes a good video. \figurename{ \ref{fig:fig1}} illustrates our proposed approach. 

We suggest a literature study of these guidelines in terms of their aspired impact on a video and its underlying characteristics (\figurename{ \ref{fig:fig1}}, $1)$). Based on these influences, we expect to identify quality characteristics that constitute good videos (\figurename{ \ref{fig:fig1}}, $2)$). In turn, the identified set of quality characteristics enables us to derive a quality model for videos (\figurename{ \ref{fig:fig1}}, $3)$). 

The resulting quality model can be used by software professionals as an orientation ($1$) to evaluate the quality of existing videos or ($2$) to guide their video production and use process in terms of pre-production, shooting, post-production, and viewing. Software professionals can produce and use videos easier at moderate costs, yet sufficient quality by knowing the quality characteristics that constitute a good video.

\section{Conclusion}
\label{sec:conclusion}
Textual specifications written in natural language are the most common medium in RE. However, this documentation option insufficiently supports requirements communication due to the inherent restrictions of available notations \cite{Al-Rawas.1996, Fricker.2010}. Different studies \cite{Lethbridge.2003, Carter.2009, Abad.2016} investigated the use of pictorial and textual representations, i.e. written specifications, as documentation option for communication. All of them concluded that there is the necessity to enrich specifications with powerful, simple, and rich documentation options in order to turn specifications into an effective means of communication.

Carter and Karatsolis \cite{Carter.2009} emphasized that when used properly including videos as a documentation option in RE could produce significant value. However, ``the challenge is to engage people with the right tools, skills, and talents to establish the context and to post-process the results properly. This suggests that research into a different set of tools aimed at [\dots] producing effective [video] clips, might be a valuable investment''\cite[p. 202]{Carter.2009}.

Many existing approaches focus on the use of videos in RE by providing corresponding processes and tools. So far, little research encountered the challenge of enabling software professionals with the necessary knowledge, skills, and talents to produce effective videos for visual communication. 

With our plan of action, we propose an approach to encounter this challenge by learning from the discipline of video production what constitutes a good video for visual communication. We strive for a quality model for videos that encapsulates this knowledge to make it easily accessible to software professionals. Such a quality model is a promising solution since it allows to (1) evaluate the quality of existing videos and (2) to guide the video production and use process. If software professionals understand what constitutes the quality of a good video, they can use this knowledge and thus the quality model as an orientation for planning, shooting, post-processing, and viewing a video. 

All in all, we conclude that videos possess a large potential as a future medium in RE. However, software professionals from research and industry currently neglect this potential due to the lack of knowledge about the challenges, actual demands, and efforts on how to communicate visually with videos. Especially, the consideration of video production as a secondary task and thus the negligence of the essential step of how to produce videos impedes their application as a medium in RE. Therefore, we assert that software professionals could enrich their communication and thus RE abilities if they knew what constitutes a good video for successful visual communication.

\section*{Acknowledgment}
This work was supported by the Deutsche Forschungsgemeinschaft (DFG) under Grant No.:~289386339, (2017 -- 2019). 

\bibliographystyle{IEEEtran}
\bibliography{IEEEabrv,references}

\end{document}